\documentclass[prl,aps,twocolumn,superscriptaddress]{revtex4-1}
\usepackage{graphicx,color}
\usepackage{amsthm}
\usepackage{amsfonts}
\usepackage{algorithmic}
\usepackage{enumerate}
\usepackage{latexsym}
\usepackage{amsmath}
\usepackage{amssymb}
\usepackage[colorlinks=true,citecolor=blue,linkcolor=blue]{hyperref}

\emergencystretch=\maxdimen
\begin{document}

\title{Zitterbewegung in time-reversal Weyl Semimetals}

\author{Tongyun Huang}
\affiliation{Department of Physics, Beijing Normal University, Beijing
100875, China\\}
\author{Tianxing Ma}
\email{txma@bnu.edu.cn}
\affiliation{Department of Physics, Beijing Normal University, Beijing
100875, China\\}
\author{Li-Gang Wang}
\email{sxwlg@yahoo.com}
\affiliation{Department of Physics, Zhejiang University, Hangzhou
310027, China\\}

\date{\today}
\begin{abstract}
We perform a systematic study of the Zitterbewegung effect of fermions, which are described by a Gaussian wave with broken spatial-inversion symmetry in a three-dimensional low-energy Weyl semimetal.
Our results show that the motion of fermions near the Weyl points is characterized by rectilinear motion and Zitterbewegung oscillation.
The ZB oscillation is affected by the width of the Gaussian wave packet, the position of the Weyl node, and the chirality and anisotropy of the fermions. By introducing  a one-dimensional cosine potential, the new generated massless fermions have lower Fermi Velocities, which results in a robust relativistic oscillation. Modulating the height and periodicity of periodic potential demonstrates that the ZB effect of fermions in the different Brillouin zones exhibits quasi-periodic behavior. These results may provide an appropriate system for probing the Zitterbewegung effect experimentally.
\end{abstract}


\maketitle

\noindent
\section{Introduction}
The Zitterbewegung (ZB) effect is characterized by extremely high-frequency oscillation which is caused by the interference between the positive and negative energy solutions of the Dirac equations. Since it was first proposed by Schr\"{o}dinger in 1930 \cite{Schro1930}, intensive studies on the ZB effect have been conducted in various fields of physics \cite{Ferrari1990,Cannata1990,Zawadzki2005,Schliemann2006,Zawadzki2006}. The aim is to realize the ZB effect experimentally, as this would be a key factor in understanding the exotic physics of relativistic quantum effects. At present, the mechanism of ZB remains mysterious because of its extremely small amplitude ($\approx{10^{-12}}$ m) and high oscillation frequency($\approx {10^{ 21}}$ Hz). Recently, the discovery of Dirac fermion materials such as graphene \cite{Novoselov2005,Zhang2005,Rusin2007,Rusin2008,Maksimova2008,Schliemann2008a,Romera2009,Garc2014,Krueckl2009,Rakhimov2011,Chaves2010}, superconductors \cite{Cserti2006} and topological insulators \cite{Shi2013} has revived the hope of detecting such elusive trembling motion, because the two interacting bands in these solids exhibit similar behavior to the Dirac equation for massless electrons in a vacuum.

Previous studies have verified that solids are a more prospective medium for observing the ZB effect than a vacuum, because the lower Fermi velocity leads to a larger period and amplitude \cite{Rusin2007,Maksimova2008}, and the transient oscillations become permanent under the special structure of the Landau levels in the presence of an external perpendicular magnetic field \cite{Rusin2008,Schliemann2008a,Romera2009,Garc2014,Krueckl2009}. Both the amplitude and frequency of oscillations attain measurable levels when the parameter of the Dirac equation is modified in those materials \cite{Rakhimov2011,Chaves2010,Cserti2006,Shi2013}, and the ZB effect has been observed in a one-dimensional ion\cite{Gerritsma2010} and a Bose-Einstein condensate\cite{LeBlanc2013}. However, this modification breaks the symmetry, and the corresponding massless electrons are unstable against perturbations because of the band-gap that opens in the two-dimensional (2D) Dirac points, As a result, those promising solids are actually non-ideal candidates for practical observation.

More recently, the Weyl semimetals (WSMs), a three-dimensional (3D) analog of graphene, have emerged as a new quantum state of matter \cite{Huang2015,Ruan2016}. Remarkably, by breaking the either time-reversal or spatial-inversion symmetry, the quadruple degeneracy of a Dirac point is broken into the double degeneracy of two Weyl points, which must appear in pairs of opposite chirality because of the fermion doubling theorem \cite{Nielsen1981,Nielsen1981a}.  The nodes of WSMs are a classic case of gapless topological bulk modes, i.e., linearly dispersive Weyl fermions that are robust and have no symmetry protection \cite{Murakami2007, Burkov2011,Burkov2011a,Wan2011}. The linearly dispersive and gapless properties imply that there should be ZB oscillations in WSMs, and the topological characteristics suggest that these oscillations will be stable near the Weyl nodes.


Apart well as being a promising possible platform for the ZB effect, chiral fermions give rise to quantum anomalies that originate from the monopole nature of the Weyl nodes \cite{Landsteiner2014,Chan2016,Lu2016}. It is natural to investigate the dynamics of Weyl fermions with different chirality, and studies of this frontier problem with broken inversion symmetry in an optical lattice have demonstrated an unusual velocity in the semimetal and a steady ZB effect in the band insulator \cite{Li2016}.

In this paper, we examine the trajectories of Weyl fermions with broken spatial-inversion symmetry in a low-energy system under a one-dimensional periodic potential.
We find that the massless Weyl fermions are generated near the Brillouin zone boundary along the direction of the potential in  reciprocal space. Their group velocities reduce to zero in the other two dimensions,
and the magnitudes of the amplitude and period of the oscillations are on the nm- and ps-scale, respectively. This is sufficiently large to provide ZB oscillations that may be observed through radiated transverse electric field\cite{Rusin2009} emitted by the trembling motion of the electron.
By using a Gaussian wave packet, we derive analytic results for the time dependence of the average displacement of Weyl fermions. Our results show that the evolution of fermions consists of rectilinear motion and ZB oscillations. By changing the parameter of the periodic potential, we also demonstrate that they depend strongly on the effective velocity.
Interestingly, the character of the Bessel function means that the maximum amplitude and period of the ZB effect exhibit a quasi-periodic behavior with the height $V_0$ and  period $L$ of the potential, and the parameter of the potential ranges across the low-value region when the fermion is away from the center of the Brillouin zone. Moreover, the motion is sensitive to the chirality of the system and the relative displacement of the fermion and the Weyl node. As a result, the two nonequivalent Weyl nodes in time-reversal WSMs are too far apart to influence the ZB effect of the fermion simultaneously.
\\

\noindent
\section{Model and method}
Compared with a time-reversal broken WSM, nonmagnetic WSMs generated by breaking the spatial-inversion symmetry could be more easily investigated using angle-resolved photoemission spectroscopy (ARPES) because the magnetic domains need not aligned\cite{Weng2016}. To date, the only means of discovering WSMs is to use ARPES to detect Fermi arcs in the surface \cite{Xu2015,Belopolski2016}. For low-energy Weyl fermions with broken inversion symmetry, the Hamiltonian for each node can be written as\cite{Grushin2016}
\begin{align}
H_0 = \sum\limits_{j = x,y,z} s\hbar v_{j}(\kappa_{j} + sb_{j})\sigma_{j}, s = \pm1 ,
 \label{eq:hamiltonian0}
\end{align}
where $v_{j}$ are the Fermi velocities in three directions, $\kappa_{j}$ are the three components of the wave vectors, $\sigma_{j}$ are the Pauli matrices, and $s = \pm1$ labels the chirality of each node. Clearly, the vector $\vec{b} = (b_{x},b_{y},b_{z})$ quantifies the separation of the two nodes in the momentum space.

The ZB effect originates from the interference between the conduction and valence bands in solid materials. One can see a the lower Fermi velocity could enhance the interference near the Dirac points or Weyl nodes. Previous studies\cite{Park2008, Brey2009} have demonstrated that the periodic potential could decrease the Fermi velocity dramatically in graphene. The cosine potential have a substantive characteristics of periodic potentia, and in theoretical study, it should be easy to have some analytical solutions and address detailed problems if a cosine potential is considered. On the other hand, the effective cosine potential, in principle, is equivalent to the square or $\delta$ periodic potential in the experiment, which is produced via electric field between electrodes\cite{Florian2015,zhangY2005}. Thus, let us assume a potential $V(x) = V_{0}\cos G_{0}x$ along the $x$-direction with periodicity $L$, where $G_{0} = 2\pi/L$, is applied to the WSM.
The Hamiltonian $H$ can be written as
 \begin{align}
 H =\sum \limits_{j = x,y,z} s\hbar v_{j}(\kappa_{j} + sb_{j})\sigma_{j} +IV(x) ,
  \label{eq:hamiltonian}
\end{align}
where $I$ is the $2\times 2$ identity matrix. It has been shown that massless Weyl fermions are generated near the Brillouin zone boundary $\vec{G}_{m} /2$ with $\vec{G}_{m} = mG_{0}\vec{x}$, and the group velocities reduce to zero in the $y$- and $z$-directions in the extreme case \cite{Park2008}. The low-energy Hamiltonian for fermions can be written as
 \begin{align}
\tilde{H}_m = \sum \limits_{j = x,y,z} s\hbar \lambda_{j}v_{j}(k_{j} + sb_{j})\sigma_{j} + I\hbar v_{x}mG_{0}/2 ,
  \label{eq:hamiltonianT}
\end{align}
with $\vec{k}\equiv\vec{\kappa}+\vec{G}_{m}/2$, $|\vec{\kappa}|\ll G_{0}$ and $\vec{\lambda} = (1, f_{m}, f_{m})$, where $f_{m} = J_{m}(\frac{2V_{0}}{\hbar v_{x}G_{0}})$ is determined by the periodic potential $V(x)$ with $J_{m}$ as the $m$th Bessel function of the first kind \cite{Brey2009}. The derivation of this expression is given in the Appendix. The difference between the Hamiltonian in Eq.~(\ref{eq:hamiltonian0}) and that in Eq.~(\ref{eq:hamiltonianT}), apart from a constant energy term, is that the velocities of fermions moving along the $y$- and $z$-directions have changed from $v_{y}$, $v_{z}$ to $f_{m}v_{y}$, $f_{m}v_{z}$, respectively. This means that fermions near $\vec{G}_{m}/2$, different from the original massless Weyl fermions in Eq.~(\ref{eq:hamiltonian0}), are massless particles with anisotropic velocities depending on the propagation direction.

\begin{figure}
\includegraphics[scale=0.5]{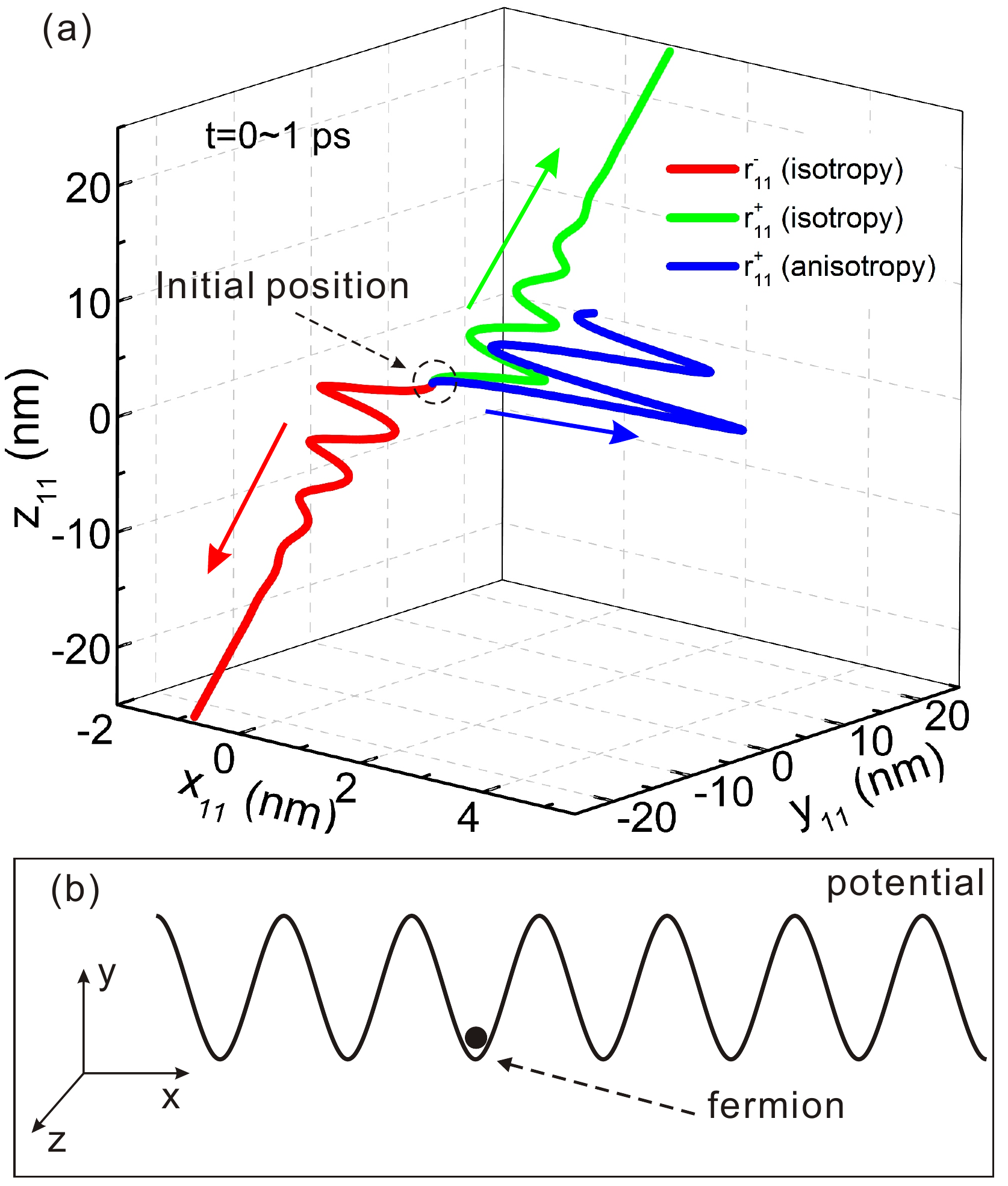}
\caption{(Color online) (a) Three-dimensional trajectories of fermions for TaAs with a cosine potential in the $x$-direction from 0-1 ps with Weyl node separation $\vec{b} = (1.735,0.025,0)\rm{\AA} ^{-1}$ and packet width $\alpha = 160 \rm{\AA}$. The green (red) line represents the motion of right-handed (left-handed) fermions in the isotropic system, and the blue line represents the motion of right-handed fermions in the anisotropic system. The initial position of the fermions was $(1.735,0.055,0.03)$, and the periodic potential factor was $f_{m} =0.1$. (b) Schematic diagram showing a Weyl fermion with a cosine potential in the $x$-direction.}
\label{Fig:1}
\end{figure}

Time-reversal WSMs have an even number of Weyl pairs, and the distance between two nonequivalent Weyl points is sufficiently far that they cannot influence the ZB effect of the fermion simultaneously. Thus, we consider only one Weyl point. The time-dependent position operator of fermions near the generated Weyl points in the Heisenberg picture $\vec{r}(t) = e^{i\tilde{H}_m t/\hbar} \vec{r}(0) e^{-i\tilde{H}_m t/\hbar}$ is a $2\times 2$ matrix. Applying the \emph{Baker-Hausdorff lemma} \cite{Sakurai1994},
\begin{align}
\vec r\left( t \right) &= \vec r\left( 0 \right) + \left( {{{it} \mathord{\left/
 {\vphantom {{it} \hbar }} \right.
 \kern-\nulldelimiterspace} \hbar }} \right)\left[ {{{\tilde H}_m},\vec r\left( 0 \right)} \right] \nonumber\\
 & + \frac{{{{\left( {{{it} \mathord{\left/
 {\vphantom {{it} \hbar }} \right.
 \kern-\nulldelimiterspace} \hbar }} \right)}^2}}}{{2!}}\left[ {{{\tilde H}_m},\left[ {{{\tilde H}_m},\vec r\left( 0 \right)} \right]} \right] +  \cdots \nonumber\\
 & = \sum\limits_{n = 0}^\infty  {\frac{{{{\left( {{{it} \mathord{\left/
 {\vphantom {{it} \hbar }} \right.
 \kern-\nulldelimiterspace} \hbar }} \right)}^n}}}{{n!}}} \left[ {{{\tilde H}_m},\left[ {{{\tilde H}_m}, \cdots \left[ {{{\tilde H}_m},\vec r\left( 0 \right)} \right] \cdots } \right]} \right],
  \label{eq:rexpland}
\end{align}
and from Eq.~(\ref{eq:hamiltonianT}), the commutator of the operators ${{\tilde H}_m}$ and $\vec r\left( t \right)$ reads
 \begin{align}
 \left[ {{{\tilde H}_m},\vec r\left( 0 \right)} \right] &= \sum\limits_{j = x,y,z} {s\hbar {\lambda _j}{v_j}\left[ { - i{\partial _j },j} \right]{\sigma _j}} {{\vec e}_j} \nonumber\\
 & =  - i\hbar \sum\limits_{j = x,y,z} {{\lambda _j}{v_j}{\sigma _j}{{\vec e}_j}},
  \label{eq:commutator}
\end{align}
where $i$ is the imaginary unit and ${\vec e}_j$ are the three components of the unit vectors.
We then obtain explicit results for the three coordinate components of $\vec{r}_{11}(t)$:
\begin{align}
x^{s}_{11}(t) &= x_{11}(0) + \frac{sf_{m}k^{'}_{x}k^{'}_{z}v^{2}_{x}v_{z}}{\omega ^{2}}t + \frac{f_{m}k^{'}_{y}v_{x}v_{y}}{2\omega ^{2}}[1 - \cos(2\omega t)] \nonumber\\
& - \frac{sf_{m}k^{'}_{x}k^{'}_{z}v^{2}_{x}v_{z}}{2\omega ^{3}}\sin(2\omega t),
 \label{eq:positionx}
\end{align}
\begin{align}
y^{s}_{11}(t) &= y_{11}(0) + \frac{sf^{3}_{m}k^{'}_{y}k^{'}_{z}v^{2}_{y}v_{z}}{\omega ^{2}}t - \frac{f_{m}k^{'}_{x}v_{x}v_{y}}{2\omega ^{2}}[1 - \cos(2\omega t)] \nonumber\\
& - \frac{sf^{3}_{m}k^{'}_{y}k^{'}_{z}v^{2}_{y}v_{z}}{2\omega ^{3}}\sin(2\omega t),
 \label{eq:positiony}
\end{align}
\begin{align}
z^{s}_{11}(t) &= z_{11}(0) + \frac{sf^{3}_{m}{k^{'}_{z}}^{2}v^{3}_{z}}{\omega ^{2}}t \nonumber\\
& + \frac{s[(f_{m}k^{'}_{x}v_{x})^{2} + (f_{m}k^{'}_{y}v_{y})^{2}]v_{z}}{2\omega ^{3}}\sin(2\omega t),
 \label{eq:positionz}
\end{align}
where the relative displacement between the fermion and the Weyl node is $k^{'}_{j} = k_{j} + sb_{j},j = x,y,z$ and the effective frequency is $\omega = \sqrt{(k^{'}_{x}v_{x})^{2} + (f_{m}k^{'}_{y}v_{y})^{2} + (f_{m}k^{'}_{z}v_{z})^{2}} $. One can see that the motion of the fermion consists of a rectilinear component and an oscillation with frequency $2\omega$. Comparing Eqs.~(\ref{eq:positionx}),~(\ref{eq:positiony}), and~(\ref{eq:positionz}), we find that the velocity in the $y$ and $z$-directions is more sensitive to the factor $f_m$ than that in the $x$-direction (see the second term of these equations), and the oscillation is influenced by the chirality of fermions (see the last term of these equations). Furthermore, the rectilinear motion is the classical velocity of the fermion.

The initial state of the fermion is described by a Gaussian wave packet \cite{Schliemann2005,Schliemann2006}
\begin{align}
\varphi (\vec{r},0) = \left(\frac{\alpha}{2\pi \sqrt{\pi}}\right)^{\frac{3}{2}}\int d^{3}ke^{-\frac{\alpha ^{2}(\vec{k}-\vec{k}_{0})^{2}}{2}}e^{i\vec{k}\vec{r}} \left(\begin{matrix}1\\0\end{matrix}\right),
\label{eq:wavepacket}
\end{align}
where $\alpha$ denotes the width of the packet, and $\vec{k}_{0}$ is the center wave vector of the packet. The unit vector $(1,0)$  is a convenient choice \cite{Schliemann2005}; the average of the $(1,1)$ component of $\vec{r}_{11}$ is written as
\begin{align}
\bar{r}_{11}(t) &= \langle\varphi(\vec{r},0)|\vec{r}_{11}(t)|\varphi(\vec{r},0)\rangle \nonumber\\
& = \frac{\alpha^{3}}{\pi^{3/2}}\int\int\int\vec{r}_{11}(t)e^{-\alpha^{2}(\vec{k} - \vec{k}_{0})^{2}}dk_{x}dk_{y}dk_{z}.
\label{eq:average}
\end{align}

For simplicity, the packet is centered at $\vec{k}_{0} = (0,0,0)$ so that there is an appropriate momentum to generate observable ZB oscillations \cite{Zhang2013} when the nodes are away from the origin of the coordinates.
\\

\noindent
\section{Results and discussion}
TaAs is a natural WSM. It belongs to the nonsymmorphic space group $I4_1md$ and has a body-centered-tetragonal structure, that lacks spatial-inversion symmetry.
The three-dimensional trajectories of TaAs fermions under a cosine potential in the $x$-direction are plotted in Fig.~\ref{Fig:1} as $\vec{b} = (1.735,0.025,0)\rm{\AA} ^{-1}$\cite{Weng2015}, the packet is centered at $\vec{k}_{0} = (1.735,0.055,0.03)$. One can see that the fermions with different chiralities (green and red lines) have opposite directions of motion, and the different chirality introduces a different phase to the system, but the trajectory of the fermion in the anisotropic situation (blue line) changes drastically. The magnitudes of the amplitude and period of the oscillations are on the nm- and ps-scale, respectively.

\begin{figure}
\includegraphics[scale=0.5]{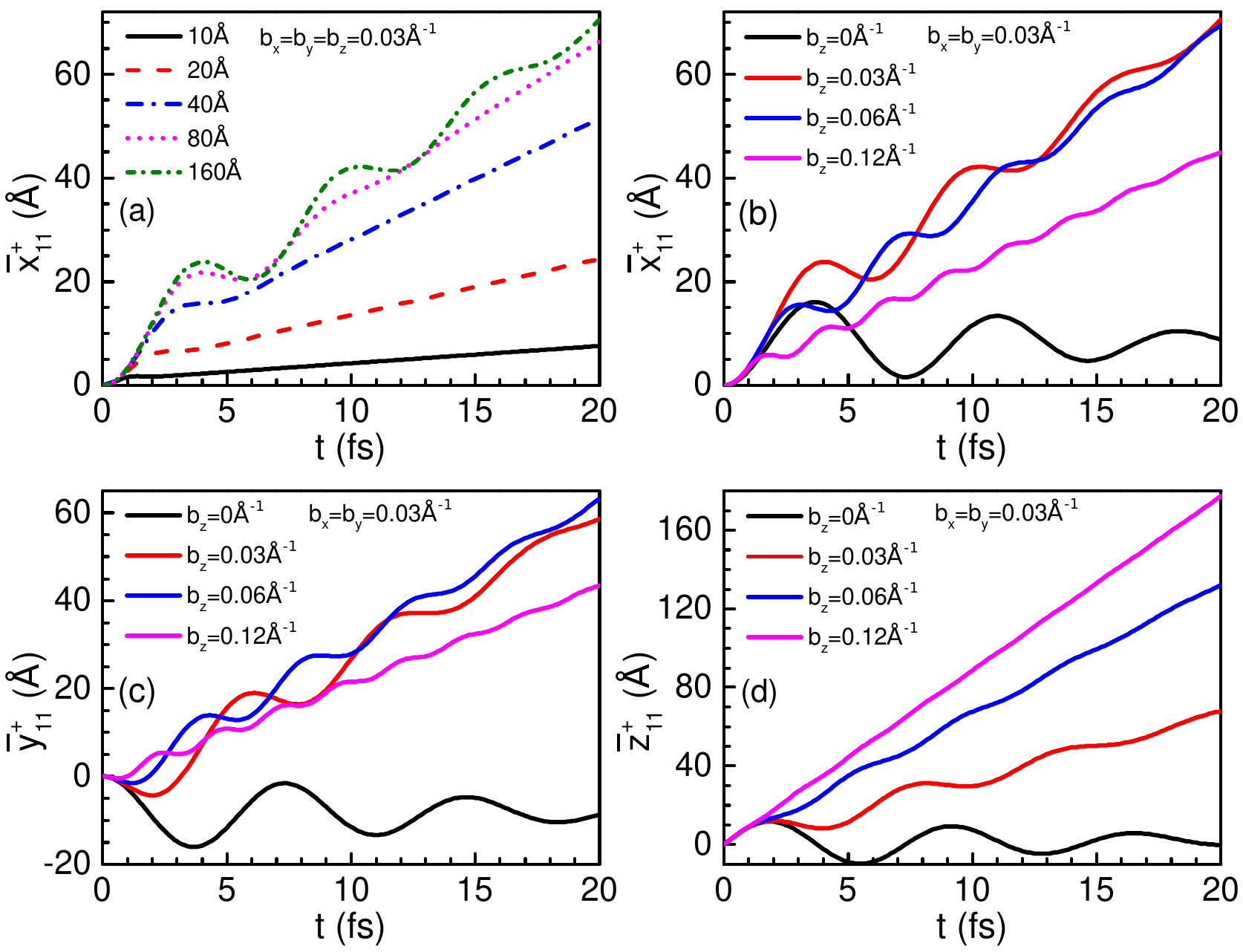}
\caption{(Color online) (a) Evolution of right-handed fermions without a periodic potential in the $x$-direction for a Gaussian wave packet of width $\alpha$ at $\vec{b} = (0.03,0.03,0.03)\rm{\AA} ^{-1}$. The average displacement of right-handed fermions for different node separations without a periodic potential is shown at $\alpha = 160 \rm{\AA}$ in the (b) $x$-direction, (c) $y$-direction, (d) $z$-direction. All Fermi velocities were $1\times 10^{8}$ cm/s.}
\label{Fig:2}
\end{figure}

To understand these unusual traits, we first study the displacement of the right-handed ($s = +1$) fermion in the $x$-direction as a function of time for a Gaussian wave packet with a different width $\alpha$ at $\vec{b} = (0.03,0.03,0.03)\rm{\AA} ^{-1}$, as shown in Fig.~\ref{Fig:2}(a). In our preliminary calculation, the Fermi velocity was set to $1\times 10^{8}$ cm/s for the isotropic system. As shown in the figure, (i) there are nearly no oscillations when $\alpha$ is small, (ii) the maximum amplitude of ZB oscillations and the velocity of the rectilinear motion in the $x$-direction increase with the width of the wave packet, whereas the period of the ZB oscillations are almost constant, and (iii) the oscillations have a transient character and may decay in several femtoseconds. These results show that the  ZB behavior is a damped oscillation with the exponential decay factor in Eq.~(\ref{eq:average}) and depends quite critically on the value of $\alpha$, consistent with previous work\cite{Rusin2007, Zhang2008}. Thus, the packet width $\alpha$ was set to $160 \rm{\AA}$ in the subsequent simulations.

Next, we focus on the relationship between the separation of two nodes and the motion of the fermion. The calculated displacements of the three coordinate components of the right-handed fermion are plotted as a function of time in Figs.~\ref{Fig:2}(b)$\sim$~\ref{Fig:2}(d), respectively. We can see that the rectilinear motion of the fermion vanishes when the node is in the $x-y$ plane ($b_{z} = 0$). This is consistent with the behavior of 2D materials \cite{Rusin2007} in which the ${\sigma_{z}}$ term is absent. The time terms of Eqs.~(\ref{eq:positionx}),~(\ref{eq:positiony}), and~(\ref{eq:positionz}), suggest that the time term is zero when $k^{'}_{z} = k_{z} + b_{z} \simeq b_{z} = 0$. Therefore, we conclude that the rectilinear motion originates from the additional momentum determined by the position of the Weyl node in the $z$-direction. In addition, the amplitude and period of the ZB effect depend sensitively on the relative displacement of the fermion and the Weyl points $k^{'}_{j}$; in other words, the smaller the value of $b_{j}$, the smaller the energy gap, leading to a lower ZB frequency and a stronger oscillating amplitude.

\begin{figure}
\includegraphics[scale=0.5]{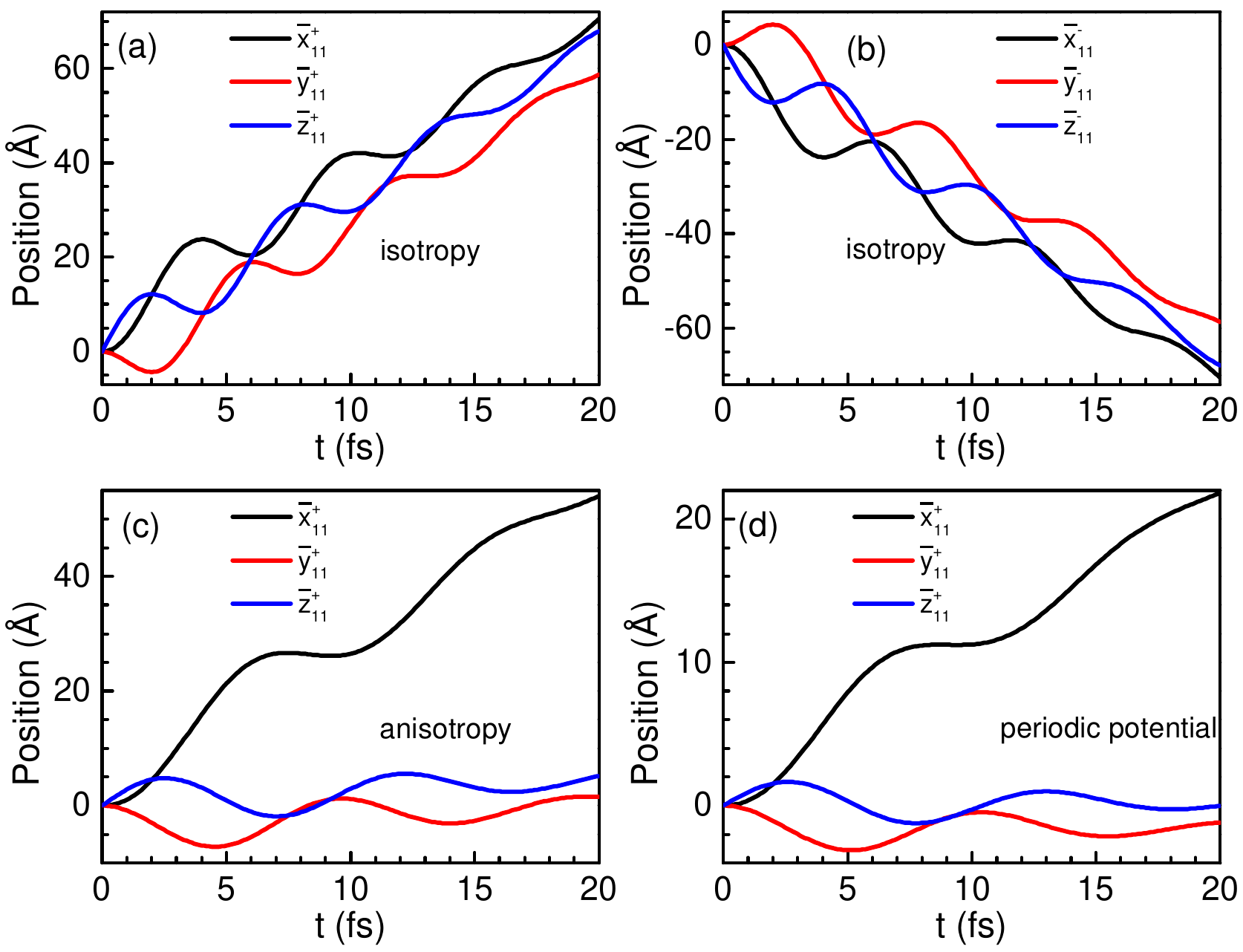}
\caption{(Color online) Average displacement of (a) (c) (d) a right-handed, and (b) a left-handed fermion in three coordinate components at $\alpha = 160 \rm{\AA}$ and $\vec{b} = (0.03,0.03,0.03)\rm{\AA} ^{-1}$; (a) and (b) are from the isotropic system, (c) is from the anisotropic system, 
and (d) is the same as (c) but with a periodic potential with $f_{m} = 0.1$. }
\label{Fig:3}
\end{figure}

From Fig.~\ref{Fig:1}, one can also find that fermions with different chiralities have opposite direction of motion. To further understand this behavior, Figs.~\ref{Fig:3}(a) and~\ref{Fig:3}(b) plot the displacement of fermions with different chiralities for the three coordinate components under the same parameters. As discussed above, the velocities of the rectilinear motion are equal and opposite since the time terms of Eqs.~(\ref{eq:positionx}),~(\ref{eq:positiony}), and~(\ref{eq:positionz}) have the chirality factor $s$. The amplitude and period of the ZB oscillations are the same for the different chiralities in all components, whereas the phase difference between the left-handed ($s =  - 1$) and the right-handed ($s =  + 1$) fermions is $\pi $ because of the opposite sign. This suggests that the chirality strongly changes both the direction of the rectilinear motion and the phase of the oscillation.

\begin{figure}
\includegraphics[scale=0.5]{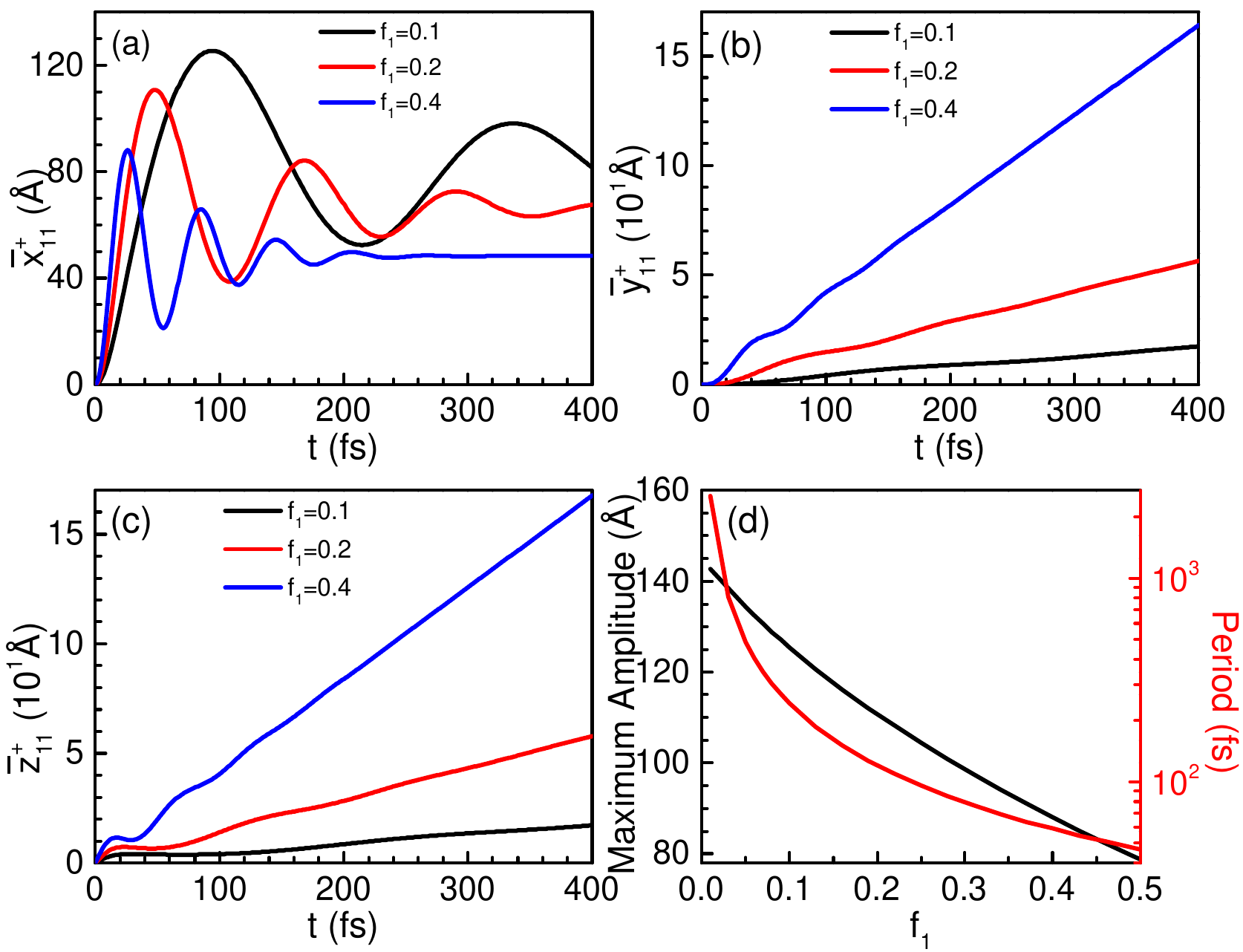}
\caption{(Color online) Evolution of right-handed fermions in the anisotropic system with periodic potentials across several $f_{1}$ values in (a) $x$-direction, (b) $y$-direction, and (c) $z$-direction. (d) Maximum amplitude and period of the ZB oscillations in the $x$-direction as a function of $f_{1}$; theWeyl points were positioned at $\vec{b} = (0,0.03,0.03)\rm{\AA} ^{-1}$.}
\label{Fig:4}
\end{figure}

The results obtained from the anisotropic case are shown in Fig.~\ref{Fig:3}(c) for the same parameters, except for $v_{x} = 1\times 10^{8}$ cm/s, $v_{y} = 3\times 10^{7}$ cm/s, and $v_{z} = 3\times 10^{7}$ cm/s \cite{Li2016}. From Figs.~\ref{Fig:3}(a) and~\ref{Fig:3}(c), it is apparent that the motion of the fermion changes dramatically. The period of the ZB oscillations for the right-handed fermion in the anisotropic system is larger than that in the isotropic system because of the small effective velocity $v_{eff} = \sqrt{v^{2}_{x} + v^{2}_{y} + v^{2}_{z}}$, which depends on the anisotropy. In addition, the change in velocity of the rectilinear motion in the $y$- and $z$-directions is more drastic than that in the $x$-direction, which is consistent with the time terms in Eqs.~(\ref{eq:positionx}),~(\ref{eq:positiony}), and~(\ref{eq:positionz}).

We now study the right-handed fermion in the anisotropic system with the periodic potential in the case of $m = 1$ (see Fig.~\ref{Fig:3}(d). Comparing Figs.~\ref{Fig:3}(c) and~\ref{Fig:3}(d), one can see that the amplitude and period of oscillations are slightly larger than that of the system without the periodic potential, which is due to the large value of $ v_x$. Therefore, we calculate the average displacement of the right-handed fermion as a function of time at $\vec{b} = (0,0.03,0.03)\rm{\AA}^{-1}$ across several $f_{1}$ values, as shown in Figs.~\ref{Fig:4}(a)$\sim$~\ref{Fig:4}(c). As $f_{1}$ decreases, the period and amplitude of the ZB oscillations increase in the $x$-direction, and the attenuation of the oscillations becomes slower. However, there is no change in the $y$- and $z$-directions, because the amplitude of oscillations is only inversely proportional to $f_{m}$ in the $x$-direction (consider Eqs.~(\ref{eq:positionx}),~(\ref{eq:positiony}), and~(\ref{eq:positionz}) with $k^{'}_{x} = k_{x} + b_{x} = 0$). In contrast, there is a positive correlation between the velocities of the rectilinear motion and $f_{1}$ in the $y$- and $z$-direction, but this vanishes in the $x$-direction. From Eqs.~(\ref{eq:positionx}),~(\ref{eq:positiony}), and~(\ref{eq:positionz}), it is apparent that the time term is zero in the $x$-direction, but proportional to $f_{m}$ in the $y$- and $z$-directions.

\begin{figure}
\includegraphics[scale=0.25]{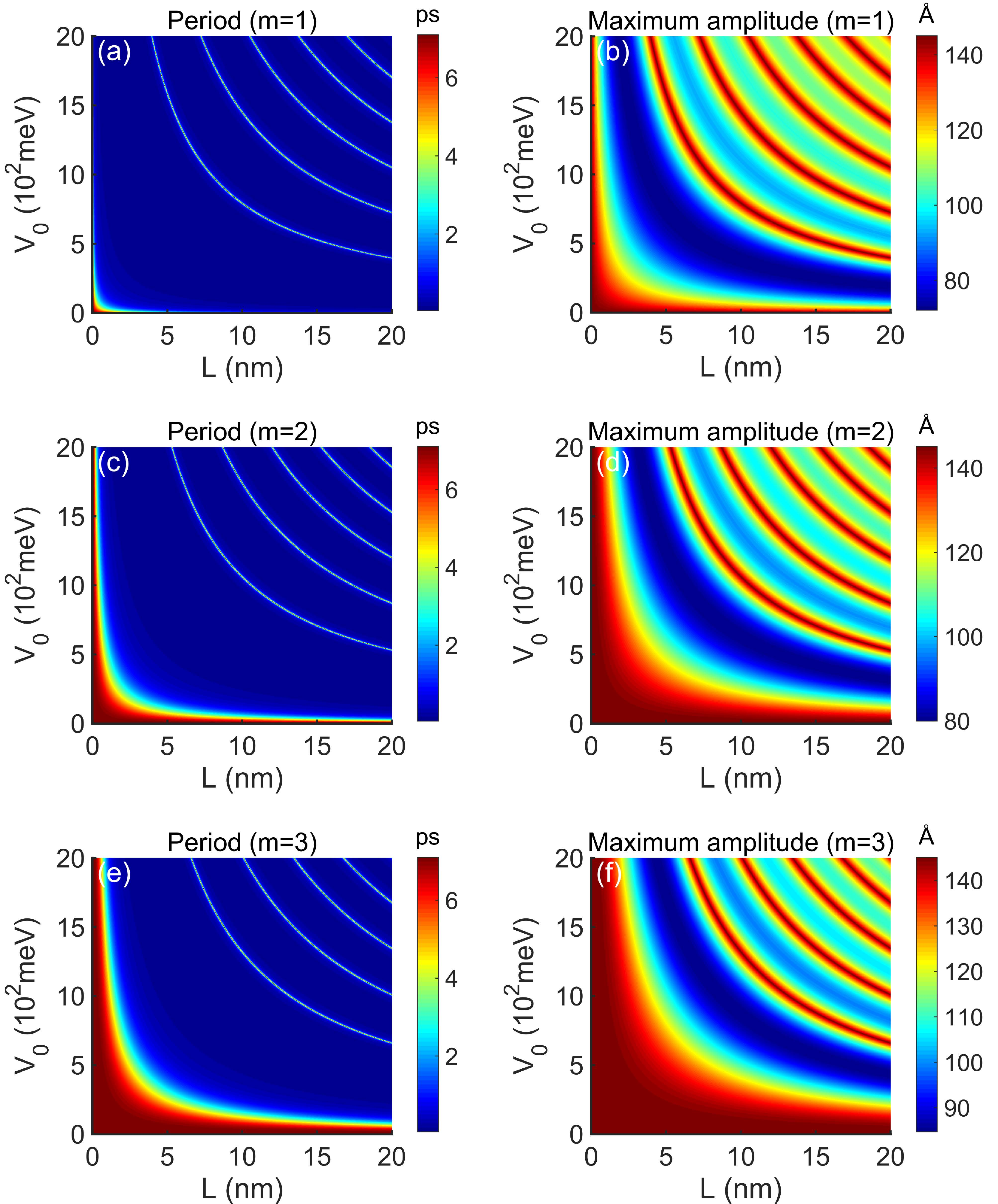}
\caption{(Color online) (a), (c), (e) Periods and (b), (d), (f) maximum amplitudes of the ZB oscillations in the $x$-direction for right-handed fermions in the anisotropic system with different cosine potential parameters $V_{0}$ and $L$. Panels correspond to different Brillouin zones: (a), (b) first Brillouin zone, (c), (d) second Brillouin zone, (e), (f) third Brillouin zone. The Weyl points were positioned at $\vec{b} = (0,0.03,0.03)\rm{\AA} ^{-1}$.}
\label{Fig:5}
\end{figure}

The long-range electron-electron interaction causes a ''logarithmic'' correction to the Fermi velocity in WSMs\cite{Isobe2012, Roy2016}. Thus, the ZB effect will show ''logarithmic'' dependence on the frequency in accordance with Eqs.~(\ref{eq:positionx}),~(\ref{eq:positiony}), and~(\ref{eq:positionz}). Similarly, the periodic potential strongly affects the ZB effect by changing the Fermi velocity. We further investigate the maximum amplitude and period of the ZB oscillations as a function of $f_{1}$ in the $x$-direction (see Fig.~\ref{Fig:4}(d)). The maximum amplitude of  the oscillations decreases from approximately $143$ to $78 \rm{\AA}$, and the period decreases from approximately 2.5 ps to 45 fs, as $f_{1}$ changes from 0.01 to 0.5. To understand the relationship between the ZB effect and the periodic potential directly, Fig.~\ref{Fig:5} shows the period and maximum amplitude of the ZB oscillations in the $x$-direction with different values of the cosine potential parameters $V_{0}$ and $L$ in the different Brillouin zones. The change in the effective velocity causes the period and the maximum amplitude of the ZB oscillations to vary quasi-periodically with the height $V_{0}$ or periodicity $L$, and their maxima decrease and become constant with increasing height $V_{0}$ or periodicity $L$. Furthermore, the period changes more drastically than the maximum amplitude because of the overlap of the trigonometric and exponential functions. From Fig.~\ref{Fig:5}, one can see that the ranges of $V_{0}$ and $L$ become broad in the low-value region, and the peaks of the period and maximum amplitude are equal when the fermion is away from the center of the Brillouin zone. All these fascinating behaviors derive from the character of the Bessel function, and our results may provide an appropriate and stable system for probing the ZB effect experimentally.
\\

\noindent
\section{Summary}
We have studied the rectilinear motion and ZB oscillation of fermions with broken spatial-inversion symmetry in a low-energy WSM. Compared with the situation in 2D materials such as graphene, the rectilinear movement has a unique character in a 3D WSMs because of the additional momentum in the $z$-direction, which allows us to identify the chiral fermion via its direction.
Furthermore, the smaller separation of the Weyl nodes results in a lower ZB frequency and a stronger oscillating amplitude, and the different chiralities of fermions in WSMs gives rise to a $\pi$ phase in the ZB oscillations.
Additionally, the effective velocity can be diminished by modulating the periodicity or height of the cosine potential in the $x$-direction when the momentum is zero in the same direction. This gives rise to a steady ZB effect that may pave the way to for investigating its behavior under perturbations.
\\

\noindent
\section{Acknowledgement} ---T. H. and T. M. were supported in part by NSCFs (grant nos. 11774033 and 11334012) and the Fundamental Research Funds for the Center Universities, grant no. ~2014KJJCB26. We also acknowledge computational support from the HSCC of Beijing Normal University. L.-G W. was supported by Zhejiang Provincial Natural Science Foundation of China under Grant No. LD18A040001, and the grant by National Key Research and Development Program of China (No. 2017YFA0304202); it was also supported by the National Natural Science Foundation of China (grants No. 11674284 and U1330203), and the Fundamental Research Funds for the Center Universities (No. 2017FZA3005).

\appendix
\section{Appendix}
This appendix presents a derivation of Eq. (\ref{eq:hamiltonianT}) in the anisotropic case, using the unitary matrix
 \begin{align}
{U_1} = \frac{1}{{\sqrt 2 }}
\left(\begin{matrix}e^{ - i\alpha \left( x \right)/2}&-e^{i\alpha \left( x \right)/2}\\e^{ - i\alpha \left( x \right)/2}&e^{i\alpha \left( x \right)/2}\end{matrix}\right),
 \label{A1}
\end{align}
where $\alpha \left( x \right)$ is written as
 \begin{align}
\alpha \left( x \right) = 2{\int_0^x {V( {{x^{'}}} )} d{x^{'}}}/\hbar {v_0},
 \label{A2}
\end{align}
with the period potential ${V( {{x^{'}}} )}$, the reduced Planck constant  $\hbar$ and the Fermi velocity $v_0$. From Eq. (\ref{eq:hamiltonian}), the Hamiltonian ${H^{'}} = U_1^ \dag H{U_1}$ reads
 \begin{align}
 H^{'}=s\hbar {v_0}
 \left(\begin{matrix}{ - i{\partial _x}} & e^{i\alpha \left( x \right)}{\left( { - {\partial _y} + i{\partial _z}} \right)}
 \\e^{i\alpha \left( x \right)}{\left( {{\partial _y} + i{\partial _z}} \right)}&{i{\partial _x}}\end{matrix}\right).
 \label{A3}
\end{align}
By applying the two pseudospin states near the  Brillouin zone boundary ($\vec{\kappa}\pm\vec{G}_{m}/2$) as basis functions, we obtain the unitary matrix
 \begin{align}
 U_2=\left(\begin{matrix}{e^{i\left(\vec{\kappa}+\vec{G}_{m}/2\right)\vec{r}}}&0
 \\0&{e^{i\left(\vec{\kappa}-\vec{G}_{m}/2\right)\vec{r}}}\end{matrix}\right).
 \label{A4}
\end{align}
After a similarity transformation ${H^{''}} = U_2^ \dag {H^{'}}{U_2}$ , the Hamiltonian ${H}$ is further given by
\begin{align}
H^{''}=s\hbar {v_0}
\left(\begin{matrix}{- i{\partial _x}+mG_{0}/2}&{\beta \left( -\partial _y+i\partial _z \right)}
\\{\beta^*\left(\partial _y+i\partial _z \right)}&{i{\partial _x}+mG_{0}/2}\end{matrix}\right),
 \label{A5}
\end{align}
where we use $- i{\partial _j }=k_j\gg {\kappa_j}\left(j=x,y,z\right) $, and $\beta = e^{-imG_{0}x+i\alpha \left( x \right)}$. Using the Fourier expansion
\begin{align}
e^{i\alpha \left( x \right)}=\sum\limits_{m =  - \infty }^\infty  {{f_m}\left( V \right)}e^{imG_0x},
 \label{A6}
\end{align}
where ${f_m}( V )$ is determined by the period potential, the matrix $M_m$ derived from the Hamiltonian $H^{''}$ can be written as
\begin{align}
M_m=s\hbar {v_0}
\left(\begin{matrix}{- i{\partial _x}+mG_{0}/2}&{f_m\left( -\partial _y+i\partial _z \right)}
\\{f_m\left(\partial _y+i\partial _z \right)}&{i{\partial _x}+mG_{0}/2}\end{matrix}\right).
 \label{A7}
\end{align}
Finally, employing a unitary transformation ${M_m^{'}} = U_3^ \dag M_m{U_3}$ with the unitary matrix
\begin{align}
U_3=\frac{1}{{\sqrt 2 }}
\left(\begin{matrix}1&1\\-1&1\end{matrix}\right),
 \label{A8}
\end{align}
the result is
\begin{align}
M_m^{'} =  s\hbar v_{0}(k_{x}\sigma_{x} + f_{m}k_{y}\sigma_{y}+f_{m}k_{z}\sigma_{z}) + I\hbar v_{0}mG_{0}/2.
 \label{A9}
\end{align}
\bibliography{reference}

\end{document}